\documentclass[aps,pra,preprint,floatfix,showpacs,12pt]{revtex4-1}
\usepackage{amsmath}
\usepackage{graphicx}
\usepackage{epsfig}
\usepackage{graphics}
\usepackage{bm}
\usepackage{amssymb}
\usepackage{psfrag}
\usepackage[T1]{fontenc}

\usepackage{color}   
\newcommand{\da}{\downarrow}
\newcommand{\ua}{\uparrow}

\begin{document}
\title{Spontaneous magnetization and anomalous Hall effect in an emergent Dice lattice}
\author{O. Dutta$^{1}$}
\email[E-mail: ]{omjyoti@gmail.com}
\author{A. Przysi\k{e}\.zna$^{1,2,3}$}
\author{J. Zakrzewski$^{1,4}$}
\affiliation{
\mbox{$^1$ Instytut Fizyki imienia Mariana Smoluchowskiego, Uniwersytet Jagiello{\'n}ski,}
 \mbox{ulica \L{}ojasiewicza 11, PL-30-348 Krak\'ow, Poland }
\mbox{$^2$ Institute of  Theoretical Physics and Astrophysics, University of Gda\'nsk, \ Wita Stwosza 57,}
\mbox{ 80-952 Gda\'nsk, Poland \hfill }
\mbox{$^3$ National Quantum Information Centre of Gda\'nsk, Andersa 27, 81-824 Sopot, Poland}
\mbox{$^4$ Mark Kac Complex Systems Research Center,
Uniwersytet Jagiello\'nski, Krak\'ow, Poland }}

\date{\today}

\begin{abstract}

Ultracold atoms in optical lattices serve as a tool to model different physical phenomena appearing originally in condensed matter. 
 To study magnetic phenomena one needs to engineer synthetic fields as atoms are neutral.
 Appropriately shaped optical potentials force atoms to mimic  charged particles moving in a given field. We present the realization of artificial gauge fields for the observation of anomalous Hall effect. Two species of attractively interacting  ultracold fermions are  considered to be trapped in a shaken two dimensional triangular lattice. 
A combination of interaction induced tunneling and shaking can result in an emergent Dice lattice. 
In such a lattice  the staggered synthetic magnetic flux appears and it can be controlled with external parameters. The obtained synthetic fields  are non-Abelian. 
Depending on the tuning of  the staggered flux we can obtain either anomalous Hall effect or its quantized version. 
Our results are reminiscent of Anomalous Hall conductivity in spin-orbit coupled ferromagnets.

\end{abstract}

\maketitle
%
\section*{Introduction}
Due to its unusual features such as quantized conductance and dissipation-less edge states, the Quantum Hall effect (QHE) \cite{Klitzing1980} has various possible applications in quantum information sciences. In practical implementation, the standard QHE
needs strong external magnetic fields and high mobility samples to occur. 
Therefore, it is particularly desirable to realize Hall effects without external magnetic fields.

In 1881,  Hall \cite{Hall1881} observed that in ferromagnetic materials there are unusually large Hall currents at low fields when compared to non-magnetic conductors \cite{Hall1879}.
Since then, theoretical explanation of this effect was a subject of a debate and it has taken a century until the physics of the phenomena were explained. This effect, known now as the anomalous Hall effect (AHE), originates from spontaneous magnetization in spin-orbit coupled ferromagnets \cite{Nagaosa, Science1, Science2}. The magnetization  breaks the time reversal symmetry while the spin-orbit coupling induces nontrivial topology of the bands \cite {Hassan,Qi}. It is not quantized for a metal, giving AHE, and quantized for insulators when Fermi energy lies in the band-gap, giving quantum anomalous Hall effect (QAHE). AHE and its quantized version can occur even in zero magnetic fields and they have been observed in various systems \cite{Qi2006, Qi2008, Liu2008, Science1, Nomura2011}.

 Haldane \cite{Haldane} in 1988 gave a theoretical proposal of an AHE without spin orbit coupling. 
He presented a quantized Hall effect without Landau levels in a system with circulating currents on a honeycomb lattice where the time reversal symmetry is broken only locally. Since then concentrated effort have been put forward to simulate AHE without the presence of a magnetic field.
The key point of such proposals is to engineer nontrivial topology of energy bands where the Hall conductance is related to the integral of the Berry curvature of the filled part of the band.
To tune the band structure in order to change its topology and induce the anomalous Hall effect, we need to create non-Abelian synthetic gauge fields  \cite{Lew, Lin2, Hauke, Sacha, Tercas}.
In the case of Ferromagnets it is done by spin-orbit coupling, in Haldane model -- by circulating currents.  In all of those proposals regarding AHE without magnetic field, one important ingredient is the presence of strong next-nearest neighbor tunneling with certain complex amplitude. Such a tunneling is in general hard to realize in normal lattices due to the exponential suppression of tunneling amplitudes with the  distance. This presents another pertinent question: is it possible to generate AHE in a lattice with only the nearest neighbor tunneling? In the present paper we present such a lattice model leading to AHE in the quantum regime. 

We show that in two-dimensional lattice combined effect of interaction induced tunneling and shaking  can induce AHE and QAHE (used in 1D, this ingredients can also lead to topological phenomena \cite{SSH}).
We focus on the systems of ultracold gases that provide versatile platform to simulate and engineer novel forms of matter \cite {Lew}. 
Our proposal consists of attractive two-species fermions (as in \cite{Dut}) trapped in a periodically shaken triangular lattice. 
Triangular lattice introduces geometrical frustration while the shaking can resonantly enhance the interaction-induced $sp$-orbital nearest neighbor tunneling. 
In effect, an emergent Dice lattice is formed accompanied by a strong staggered flux which, due to inclusion of $p$-orbitals, leads to a formation of synthetic non-Abelian fields. 
The system shows spontaneous magnetization accompanied by appearance of anomalous Hall conductivity forming an ultracold gas analogue of spin-orbit coupled ferromagnetic insulators. 
Furthermore, we show that, in presence of a strong staggered field, one can reach the regime of quantized Hall conductivity. 
This is a proposal of an experimentally realizable system with AHE without spin-orbit coupling.

\section*{Results}
\subsection*{The model} 

Consider an unequal mixture of two-species attractive ultracold fermions (denoted by $\uparrow,\downarrow$) trapped in a triangular lattice with fillings $n^{\ua}=1/3$ and $n^{\da}>1/3$. A strong attractive contact interaction between atoms leads to pairing -- formation of composites between the $\uparrow$ and $\downarrow$-fermions, as studied experimentally for different lattice geometries  \cite{Chin, Strohmaier, Hacker}. 
We define a composite creation operator $\hat{c}^{\dagger}_{\mathbf{i}}=\hat{s}^{\dagger}_{\uparrow \mathbf{i}}\hat{s}^{\dagger}_{\downarrow \mathbf{i}}$ with the corresponding 
 number operator $\hat{n}^c_{\mathbf{i}}=\hat{c}^{\dagger}_{\mathbf{i}}\hat{c}^{}_{\mathbf{i}}$.  
$\hat{s}^{\dagger}_{\sigma \mathbf{i}}, \hat{s}_{\sigma \mathbf{i}}$ are the creation and annihilation operators of the $\sigma$ fermions in the respective $s$-bands. { The composites are hardcore bosons
which anti-commute at the same site, $\{\hat{c}_{\mathbf{i}}, \hat{c}^{\dagger}_{\mathbf{i}} \} =1$,
and commute for different sites, $\left[\hat{c}_{\mathbf{i}}, \hat{c}^{\dagger}_{\mathbf{j}} \right]=0$ for $\mathbf{i} \neq \mathbf{j}$ \cite{Mic}. }

{
We consider three lowest bands of the triangular two-dimensional (2D) lattice (we assume some tight trap in the third direction as in typical 2D cold atoms experiments  \cite{Seng,Struck}). For sufficiently deep optical lattices the structure of the bands may be understood using a harmonic approximation for separate sites. The lowest band
is the $s$-band with two close in energy $p$-orbitals forming the excited bands. Typically fermions (for low filling) reside in the $s$-band. However, once the composite occupies a given site an additional fermion coming to this site must land in the excited band due to the Pauli exclusion principle.}

{
The harmonic approximation typically underestimates the tunneling coefficients (for a discussion see a recent review \cite{Dutta2015}). This is of no importance for the following since we assume that by using the well developed 
lattice shaking techniques, one can} tune the standard intra-band tunneling to negligible values \cite{Andre, Seng,Struck}. Such a shaking simultaneously makes the intra-band interaction induced tunneling \cite{Hirsch89,Best09,Mering11,Dutta11,Luehmann12,Lacki13} (called also bond-charge tunneling) vanishingly small.  The only remaining tunneling mechanism is then the $sp$-inter-band interaction induced tunneling \cite{Dut} which can be resonantly enhanced  adjusting the shaking frequency {(note that the standard single-body tunneling between $sp$ orbitals vanishes in Wannier function representation).}
{Therefore, the system at low-energies consists of the composites and the excess $\da$-fermions with filling $n^{\da}-n^{\ua}$. Note that the $\da$- or $\ua$-fermions of the composites cannot undergo $sp$-tunneling without breaking the strong pairing - which costs energy. {Similarly, as discussed in detail in \cite{Mic,Strohmaier}, the tunneling of the composites to a neighboring vacant site as a whole is extremely small, (see Methods section) so it is neglected}.  
Thus, the low-energy local Hilbert subspace is spanned   by $\hat{c}^{\dagger}_{\mathbf{i}}|0\rangle, \hat{p}^{\dagger}_{\pm, \mathbf{i}}\hat{c}^{\dagger}_{\mathbf{i}}|0\rangle$,  and $ \hat{s}^{\dagger}_{\mathbf{i}}|0\rangle $ states, where 
$\hat{s}_{\mathbf{i}}, \hat{p}_{\pm, \mathbf{i}}$ denote the excess $\downarrow$-fermions operators in the $s$- and $p$-orbitals. The latter are written in the chiral representation $\hat{p}_{\pm}=(\hat{p}_x \pm i\hat{p}_y)/\sqrt{2}$.} {Within this subspace, one can show that the composite number operator equals  the $\uparrow$-fermions number operator, $\hat{n}^c_{\mathbf{i}}=\hat{n}^{\ua}_{\mathbf{i}}$ and  the densities $n^c=n^{\ua}=1/3$. Other important relations are: i) 
{$\hat{s}^{\dagger}_{\mathbf{i}}\hat{c}^{\dagger}_{\mathbf{i}}=0$} --  a composite and an excess $s$-fermion cannot occupy the same site due to the Pauli-exclusion principle; ii) $[\hat{n}^c_{\mathbf{i}},\hat{p}_{\pm, \mathbf{j}}=0]$, and iii)  $[\hat{n}^c_{\mathbf{i}},\hat{s}_{\mathbf{j}}]=0$ for $\mathbf{i} \neq \mathbf{j}$. }

The effective Hamiltonian for the composites and the excess $\downarrow$-fermions consists of three parts (see Methods for more details {and Fig.~\ref{sup_fig} for visualization of tunnelings}):
 $H_{sp}$ describing interaction-induced $sp$-tunneling,  $H_{onsite}$ describing energies and local contact interactions, and $H_{shaking}$ describing the driving force. First of them reads,
\begin{equation}\label{sptun}
\frac{H_{\rm sp}}{J_{\rm sp}}= \frac{1}{\sqrt{2}}\sum_{\mathbf{i},\bm{\delta},\sigma=\pm} f_{\bm{\delta}\sigma} \hat{p}^{\dagger}_{\sigma,\mathbf{i}}\hat{n}^c_{\mathbf{i}} \hat{s}_{\mathbf{i}+\bm{\delta}} + {\it h.c},
\end{equation}
where vectors connecting nearest-neighbors  in the triangle lattice are $\bm{\delta}=\pm\bm{\delta}_0, \pm\bm{\delta}_{\pm}$with $\bm{\delta}_0=(1,0),  \bm{\delta}_{+}=(1/2,\sqrt{3}/{2}),  \bm{\delta}_{-}=(1/2,-\sqrt{3}/{2})$. Due to the angles created by the different $\bm{\delta}$ vectors, in the chiral representation an additional phase factor $f_{\bm{\delta}}$ appears. 
In the harmonic approximation of the triangular lattice potential, this phase factor is given by $f_{\bm{\delta}\sigma} = \exp\left[-i\sigma\tan^{-1}\left(\delta_y/\delta_x\right)\right]$. The tunneling $J_{sp}$ is given in terms of the $s$- and $p$-band Wannier functions $W_\mathbf{i}^{00}(x,y)$ and $W_\mathbf{i}^{10}(x,y)$  as 
\begin{equation} 
J_{sp}=g_{\rm 2D}\int\int W_\mathbf{i}^{10}(x,y)(W_\mathbf{i}^{00}(x,y))^2W_\mathbf{i+\mathbf{\delta}_0}^{00}(x,y)dx dy,
\end{equation}
with the contact interaction strength $g_{\rm 2D}$ adjusted for a quasi-2D geometry (with a tight harmonic confinement along $z$) \cite{Petrov1}. The second part {gives the on-site  Hamiltonian including higher band energy contribution and contact interactions.} It reads \cite{Dut},
\begin{eqnarray}\label{interaction}
H_{\rm onsite} &=& {U_2}\sum_\mathbf{i} \hat{n}^c_i  
+ U_{sp} \sum_{\mathbf{i},\sigma=\pm} \hat{n}^c_\mathbf{i}\hat{n}^{}_{\sigma\mathbf{i}}
+ E_1\sum_{\mathbf{i},\sigma=\pm} \hat{n}^{}_{\sigma\mathbf{i}}, 
\end{eqnarray}
where $U_2$ denotes the energy of the composites and $U_{sp}$ is the additional interaction energy to occupy the $p$-orbital of a composite filled site. $E_1$ is the single-particle excitation energy of the $p$-band. Shaking with elliptical periodic driving force leads to \cite{Seng},
\begin{equation}
H_{\rm shaking}=\sum_{\mathbf{i}} \mathbf{i}\cdot \mathbf{F}_t (\hat{n}^{}_{s\mathbf{i}}+\hat{n}^{}_{+\mathbf{i}}+\hat{n}^{}_{-\mathbf{i}}),
\end{equation}
with the shaking force $\mathbf{F}_t=[-K_1\sin(\Omega t)\hat{x}+K_2\cos(\Omega t+\Phi) \hat{y}]$. We consider the case where $J_{sp} \ll U_2, U_{sp} \leq \Omega$. That allows us to use rotating-wave approximation and Floquet theorem and to average terms fast oscillating in time (see Methods). 

The $sp$-tunneling will be resonantly enhanced when the energy to occupy the $p$-bands is  an integer multiple of the shaking frequency. 
This translates into the condition that $E_1+U_{sp}=N\Omega$ for integer $N$. 
The resonance order, $N$, can be controlled by varying either the lattice depth, interaction strength or the driving frequency. 
The time-averaged Hamiltonian then becomes,
\begin{eqnarray}\label{timeavgsp}
\frac{H_{\rm avg}}{J_{\rm sp}} &=& \frac{1}{\sqrt{2}} \sum_{\mathbf{i},\mathbf{\delta},\sigma=\pm} F_{\bm{\delta}} f_{\bm{\delta}\sigma} \hat{p}^{\dagger}_{\sigma,\mathbf{i}}\hat{n}^c_{\mathbf{i}} \hat{s}_{\mathbf{i+\delta}} ,\end{eqnarray}
where 
\begin{eqnarray}\label{tavg}
F_{\bm{\delta}} &=& \frac{1}{2\pi}\int^{2\pi}_0 \exp\left[iN\Omega t-i K_{\mathbf{\delta}}\cos(t+\alpha_\mathbf{\delta})\right]dt 
= \mathcal{J}_N(K_{\mathbf{\delta}}/\Omega)\exp[-i N \alpha_{\mathbf{\delta}}],
\end{eqnarray} 
with $\mathcal{J}_N(x)$ being the Bessel function of the first kind with integer order $N$.  The amplitudes are
$K_{\mathbf{\delta}_{0}}=K_1 $ and $K_{\mathbf{\delta}_{\pm}}=\left[K^2_1+3K^2_2 \pm 2\sqrt{3}K_1K_2 \sin\Phi \right]^{1/2}$. The phase factor $\alpha_{\bm{\delta}}=0$ for $\bm{\delta}=\pm\bm{\delta}_0$ and $\alpha_{\bm{\delta}}=\tan^{-1}\left[ \frac{\sqrt{3}\cos\Phi}{1\pm\sqrt{3}\sin\Phi} \right]$ for $\bm{\delta}=\pm\bm{\delta}_\pm $. {The effective tunneling strength  may be characterized by $J'_{\rm sp}(N,\delta) = J_{\rm sp}\mathcal{J}_N(K_{\mathbf{\delta}}/\Omega)$.}
Moreover, lattice shaking also induces phases to the $sp$-tunnelings \eqref{tavg} as illustrated in Fig.\ref{fig0}(a) (see also Methods).


\subsection*{The ground state composite structure}
 The composite number operator $\hat{n}^c_{\mathbf{i}}$ commutes with the Hamiltonian \eqref{timeavgsp}, $[\hat{n}^c_{\mathbf{i}}, \hat{H}_{\rm avg}]=0$. Therefore, we can characterize a site by the presence or the absence of a composite, i.e. $n^c_{\mathbf{i}}=1,0$  which makes the Hamiltonian, \eqref{timeavgsp}, quadratic in operators {\it for a particular realization} of composite configuration. 
{For a given composite configuration, we then diagonalize the quadratic Hamiltonian and fill up the energy levels depending on the excess $\downarrow$-fermions filling $n^{\downarrow}-n^{\uparrow} \leq 1/3$.  We find the ground state composite structure by comparing the energies of different composite configurations using simulated annealing on $6\times6$ up to $20\times12$ lattices with periodic boundary conditions. For details about the parameters of simulated annealing, we refer to Ref. \cite{Dut}. The resulting ground state self-organized structure of the composites resembles a Dice lattice and is shown in Fig.~\ref{fig0}(a). Its basis consists of three sites denoted A, B and C. 
The A site consists of two orbitals $p_{+}$ and $p_{-}$ whereas the site $B$ and $C$  have only $s$-orbitals. The basis vectors for the Dice lattice are given by $\mathbf{a}_1=(3/2,\sqrt{3}/2)$ and $\mathbf{a}_2=(3/2,-\sqrt{3}/2)$. For any deviation from the $1/3$ filling of the composites, the excess composites or vacancies will show up as impurities on top of the Dice lattice as long as the density of such impurities is small ($n_{\rm imp} \ll 1/3$).}

{ To understand the origin of the Dice structure}, consider first a composite at some chosen site A. The  energy is minimized when all the neighboring sites (forming hexagon with the site A in the center) are without composites. This facilitates the $sp$ tunneling from a $p$ orbital at site A to the neighboring sites.
Any composite on these neighboring sites increases the energy by $J_{\rm sp}F_{\rm avg,\mathbf{\delta}}$. Thus, for a composite filling of $n^c=1/3$, the delocalization area is maximized by filling the lattice with hexagons with a composite at their center. 

{Assuming the ground state configuration is fixed, the effective Hamiltonian for the excess fermions is quadratic and thus easily diagonalized to yield the band structure. The behavior of the excess fermions is then that of an ideal Fermi gas with such band structure, which is easily computed.}

\subsection*{Creation of the staggered field} 

\begin{figure}
\includegraphics[width=140mm,natwidth=610,natheight=642]{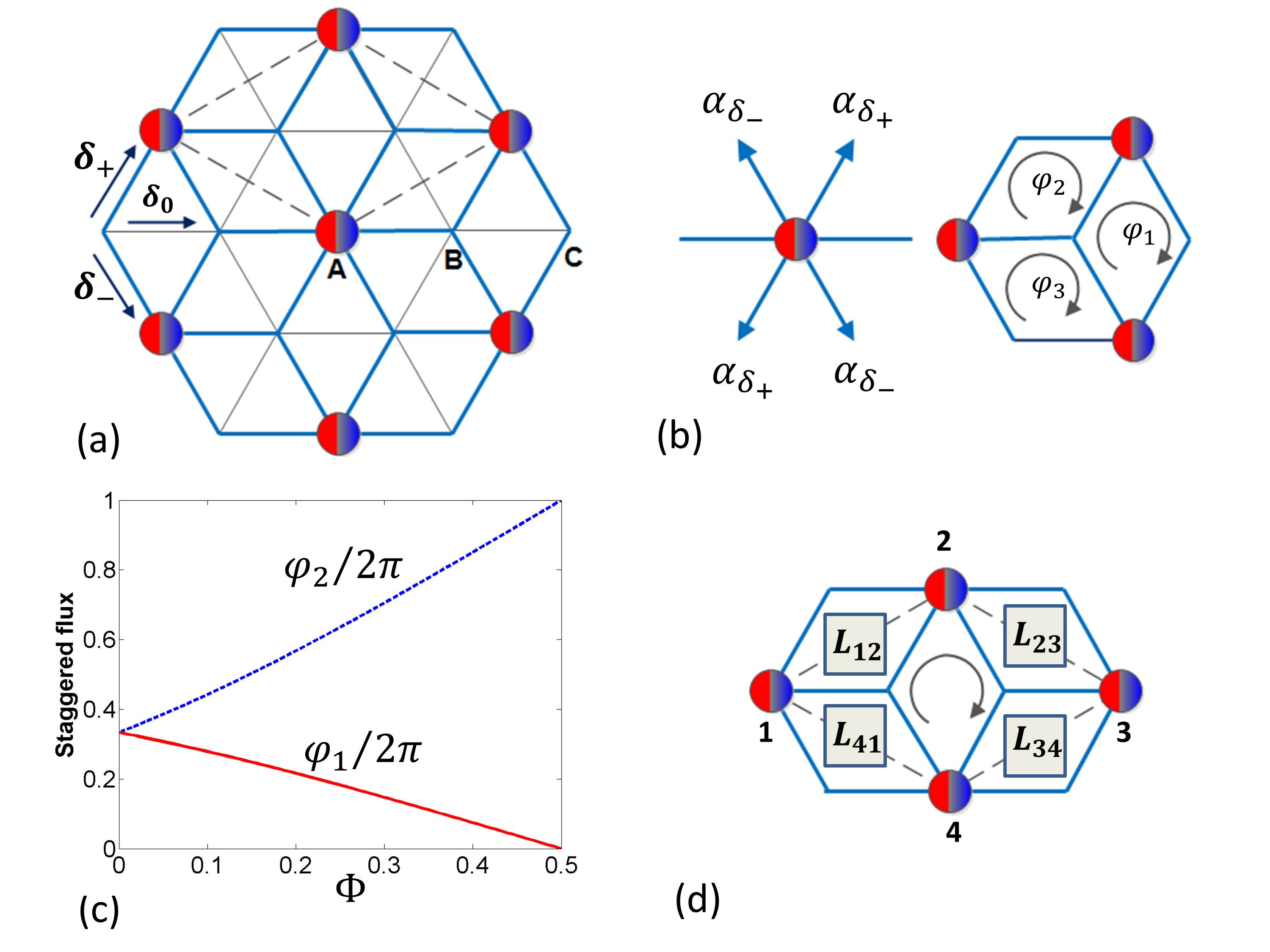}
\caption{\label{fig0} (a) The representation of the system considered in the paper. Red-and-blue spheres refer to the composites. The thin lines denote the bonds in the original triangular lattice and the blue lines represent the bonds for the excess $\downarrow$-fermions in the emerged dice lattice. 
On the composite occupied sites we have $p$-orbitals for the excess fermions while on vacant sites we have $s$-orbitals for the excess fermions. 
$\bm{\delta}_0,\bm{\delta}_{\pm}$ correspond to the vectors connecting the nearest neighbors and the sites $\mathbf{A}, \mathbf{B},\mathbf{C}$ symbolize the basis for the Dice lattice. (b) Left panel: The tunneling phases described in Eq.\eqref{tavg}. Right panel: Three elementary cells present in the dice lattice and the corresponding fluxes in those cells. The arrows show the direction along which the fluxes are calculated. (c) The magnitude of the fluxes in each cell  plotted as a function of the shaking phase $\Phi$ (in the units of $\pi$). (d) The elementary plaquette for the $A$ sites in the dice lattice along with the matrices ${L}_{ll+1}$ coupling the sites $l$ and $l+1$.}
\end{figure}

{For the Dice lattice considered here, one can construct two kind of plaquettes: i) The three plaquettes as shown in Fig.~\ref{fig0}(b) where the particle traverses the closed path involving $p_x \leftrightarrow s \leftrightarrow p_x \leftrightarrow s$ orbitals. Such a path does not mix the $p_x$ and $p_y$ orbitals. Due to the phases of the tunneling amplitudes, a particle
going through each of those plaquettes (denoted by $\varphi_1,\varphi_2$ and $\varphi_3$, see Fig.\ref{fig0}(b) and calculated along the direction of the arrow) acquire fluxes due to Aharonov-Bohm effect. We find that the induced flux is staggered in nature as the phases obey the constraint $mod(\varphi_1+\varphi_2+\varphi_3,2\pi)=0$. These fluxes are calculated by taking into account a single p-orbital.  
Fig.~\ref{fig0}(c) shows the flux strengths for $N=1$. In particular, for $\Phi=0$, $ \varphi_1=\varphi_2=\varphi_3=2\pi/3$ which is equivalent to a uniform magnetic flux of the same magnitude. With growing $\Phi$, fluxes change with all fluxes vanishing at $\Phi=\pi/2$.}


{ ii) The other kind of plaquette involves the $A$ sites containing $p_x,p_y$ orbitals as given by the parallelogram shown in Fig. \ref{fig0}(d). A particle going around such plaquette picks up a non-Abelian flux. Consider the transport from site ``1'' to ``2''. The process can go either via the upper or the lower path with two consecutive $sp$ tunnelings with strengths  $J'_{\rm sp}(N,\delta_0) = J_{\rm sp}\mathcal{J}_N(K_{\mathbf{\delta_0}}/\Omega)$ and
$J'_{\rm sp}(N,\delta_+) = J_{\rm sp}\mathcal{J}_N(K_{\mathbf{\delta_+}}/\Omega)$. The effective amplitude becomes ${\cal T}_{12}=J'_{\rm sp}(N,\delta_0)J'_{\rm sp}(N,\delta_+)/2$. The  kinetic energy term around the plaquette for $A$ sites may be written as}
{
\begin{equation}
H_{\rm kin,A}=\Psi^\dagger_2 {\cal T}_{12}{L}_{12}\Psi_1 + \Psi^\dagger_3  {\cal T}_{23}{L}_{23}\Psi_2 + \Psi^\dagger_4  {\cal T}_{34}{L}_{34}\Psi_3 + \Psi^\dagger_1  {\cal T}_{41}{L}_{41}\Psi_4 + h.c,
\end{equation}}
where the array $\Psi_l=(p_+, p_-)_l$ denotes  $p$-orbitals at site $l$. 
The corresponding link variables connecting the neighboring $A$ sites along the clockwise direction are given by $2\times 2$ matrix ${L}_{mn}$ with
\begin{equation} \label{Lmat}
  L_{12}=\begin{pmatrix}
    \cos\left(\pi/3+\alpha_{\bm{\delta}_{+}}\right) & e^{i\pi/3}\cos\alpha_{\bm{\delta}_{+}}\\
    e^{-i\pi/3}\cos\alpha_{\bm{\delta}_{+}}& \cos\left(\pi/3-\alpha_{\bm{\delta}_{+}}\right)
  \end{pmatrix}
  .
\end{equation}
$L_{23}$ {(${\cal T}_{23}$)} is given by changing $\pi/3 \rightarrow - \pi/3$ and $\bm{\delta}_{+}\rightarrow \bm{\delta}_{-}$ in the expression for $L_{12}$ {(${\cal T}_{23}$)}. Moreover, we find that $L_{12}=L_{34}$ and $L_{23}=L_{41}$ {with similar relations for ${\cal T}_{ij}$)} and they depend on the staggered flux through the phase of the tunneling amplitudes. The link variables are not unitary, which makes it not straightforward to describe them as synthetic non-Abelian fields. Nonetheless one can polar decompose them {to, ${L}_{mn}=\mathcal{S}_{mn}\mathcal{U}_{mn}$, where 
$\mathcal{U}^{\dagger}_{mn}\mathcal{U}_{mn}=1$. Such decompositions are possible as the $L_{nm}$ matrices are positive semi-definite.} Then one can define a corresponding Wilson loop parameter \cite{Linear}
\begin{equation}
W=Tr\left[\mathcal{U}_{12}\mathcal{U}_{23}\mathcal{U}_{34}\mathcal{U}_{41}\right].
\end{equation} 
The Wilson loop parameter has (i) an intrinsic contribution {(appearing at $\Phi=\pi/2$, with no staggered flux through an individual plaquete in 
Fig.~\ref{fig0}(b)} due to the
appearance of the $sp$-band tunneling and (ii) an extrinsic contribution due to the external staggered flux induced by shaking.  As a result, the link matrices are of non-Abelian nature ($W \neq 2$ ) for any shaking phase $\Phi$. 

\subsection*{Spontaneous magnetization}
\begin{figure}
\includegraphics[width=140mm,natwidth=610,natheight=642]{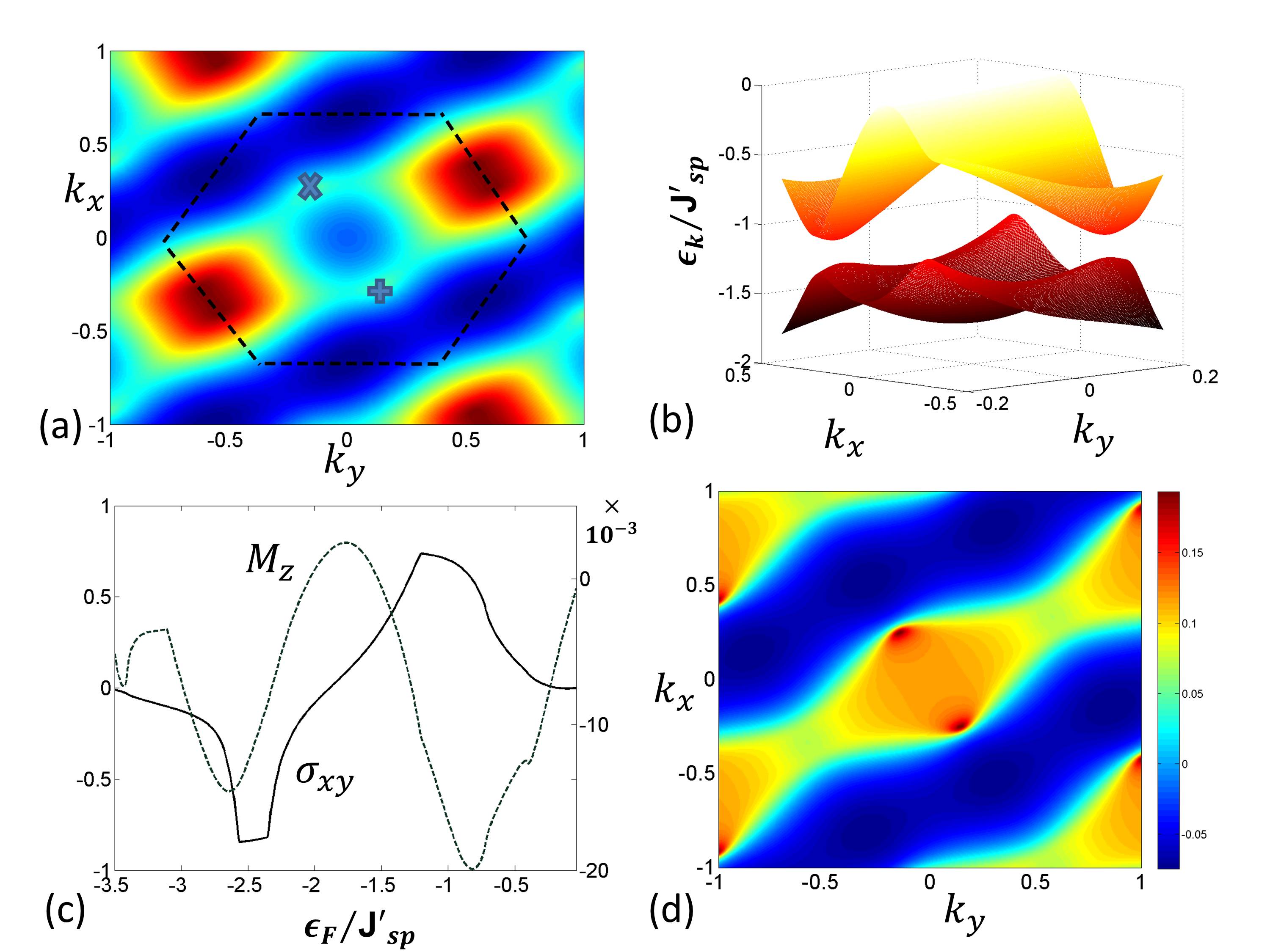}
\caption{\label{fig1} 
(a) The dispersion relation for the lowest energy band as a function of a lattice momentum $\mathbf{k}$ for zero staggered flux, $\varphi_1=\varphi_2=\varphi_3=0$ corresponding to shaking phase 
$\Phi=\pi/2$ and $N=1$. The dark blue part denotes low energy regions. The $\times$ and $+$ denote positions of  Dirac points. (b) The dispersion $\epsilon_{\mathbf{k}}$ for the first two band in the presence of small staggered flux for $\Phi=\pi/4$. The presence of the staggered flux along with the non-Abelian nature of the system helps to open a gap near the Dirac points. The tunneling strength is given by $J'_{sp}=J_{sp}J_1(K_1/\Omega)$. (c) The magnetization \eqref{mag} and Hall conductivity as a function of Fermi energy for $\Phi=\pi/4$ and $N=1$. Spontaneous magnetization appears due to  time-reversal symmetry breaking. {The Berry curvature has a non-zero contribution near the band-touching points. Contribution of local Berry curvature from such band-touching points} results in a finite Hall conductivity which shows plateau like structure due to the presence of the gap between the first two 
bands. (d) The magnetization $\mathcal{M}_{\mathbf{k}}$ in momentum space (from Eq.~\eqref{mag}) as a function of crystal momentum $\mathbf{k}$ shows sharp peaks near the two Dirac points for $\Phi=\pi/4$ and $N=1$. They correspond to the presence of monopole like structure in the corresponding Berry phase. }
\end{figure}

 First, we study the  behavior of the system in the absence of staggered flux realized for $N=1, \Phi=\pi/2$ Fig.~(\ref{fig0}). The effective non-Abelian field is intrinsic in nature and the corresponding dispersion relation for the lowest energy band is shown in Fig.~\ref{fig1}a. The main characteristic of the dispersion relation is the appearance of two non-equivalent Dirac cones and disappearance of flat bands. This is in contrast to the dispersion relation in a normal Dice lattice {where the dispersion relation contains an intersecting Dirac cone and a flat band.} 
Moreover, above a certain Fermi energy (of the excess fermions), the first two bands are degenerate.
When we  introduce the staggered flux,  the dispersion  changes and the  gap opens at the band touching points (Fig.~\ref{fig1}b). 
Once the Fermi energy is higher than the gap, the two bands become degenerate again. This is in a stark contrast to other situations with nearest neighbor tunneling where staggered flux leads only to the movement of the Dirac cones \cite{Tarruell, Lim1} and
to opening a gap one either needs long-distance tunneling \cite{Haldane, Hauke}, uniform magnetic field or synthetic non-Abelian fields along with magnetic field \cite{Kubasiak}.

\subsection*{Anomalous Hall effect} 
Consider the local magnetization ($\mathcal{M}_z$) in position space as well as magnetization in momentum space defined as 
\begin{eqnarray}\label{mag}
\mathcal{M}_z&=&\langle \hat{n}_{\mathbf{i}+}\rangle - \langle \hat{n}_{\mathbf{i}-}\rangle, \nonumber\\
\mathcal{M}_{\mathbf{k}}&=&\langle \hat{n}_{\mathbf{k}+}\rangle - \langle \hat{n}_{\mathbf{k}-}\rangle.
\end{eqnarray}
A non-zero local magnetization characterizes the breaking of time-reversal symmetry as the particles acquire local angular momentum due to the particle number difference between the $p_{+}$ and $p_{-}$ orbitals. 
First, we find that the  presence of non-zero staggered flux immediately results in non-zero $\mathcal{M}_z$ and in opening of the gap. 
Thus, appearance of non-zero $\mathcal{M}_z$ can be used as an indirect evidence for the presence of a gap in our system. 
The local magnetization is shown in Fig.\ref{fig1}c (dashed line)
for a small staggered flux. It vanishes only when the first two bands are totally filled. The presence of spontaneous magnetization (spontaneous time-reversal symmetry
breaking) is reminiscent of spin-orbit coupled ferromagnets \cite{Nagaosa}. 
Moreover, to look into the topological nature of the system, we define the intrinsic Hall conductivity,
\begin{equation}
\sigma_{xy}=\sum_{\epsilon_n<\epsilon_F} \Omega_n(\mathbf{k})/2\pi.
\end{equation}
The Berry curvature, $ \Omega_n(\mathbf{k})$,  for the $n$-th band is given 
by $ \Omega_n(\mathbf{k}) = \bm{\nabla}_{\mathbf{k}} \bm{\times} \left\langle {u}_{n\bm{q}}\left\|\bm{\nabla}_{\mathbf{k}}\right \|{u}_{n\bm{q}} \right \rangle $ 
where $|{u}_{n\bm{q}} \rangle $ denotes an eigenvector for the $n$-th band. The total Hall conductivity $\sigma_{xy}$ then depends on the Fermi energy $\epsilon_F$ of the system as shown in [Fig.\ref{fig1}c (solid line)].
We find that the local Berry curvature is concentrated near the Dirac points which results in a non-zero contribution to $\sigma^n_{xy}$ when $\epsilon_F$ is in the band.  
As $\epsilon_F$ enters the band gap, we find that $\sigma^n_{xy}$ flattens at a value $>1/2$. 
This can be ascribed to the presence of two Dirac cones near the band gap. As we increase $\epsilon_F$, the contribution from the next band begins to play a role and eventually the conductivity changes sign. The second peak appears when the Fermi energy reaches the maximum of the first band. Such structures in conductivity have been predicted to arise due to the presence of magnetic monopoles in the momentum space \cite{Nagaosa1}.

\subsection*{Quantum Anomalous Hall effect}
\begin{figure}
\includegraphics[width=140mm,natwidth=610,natheight=642]{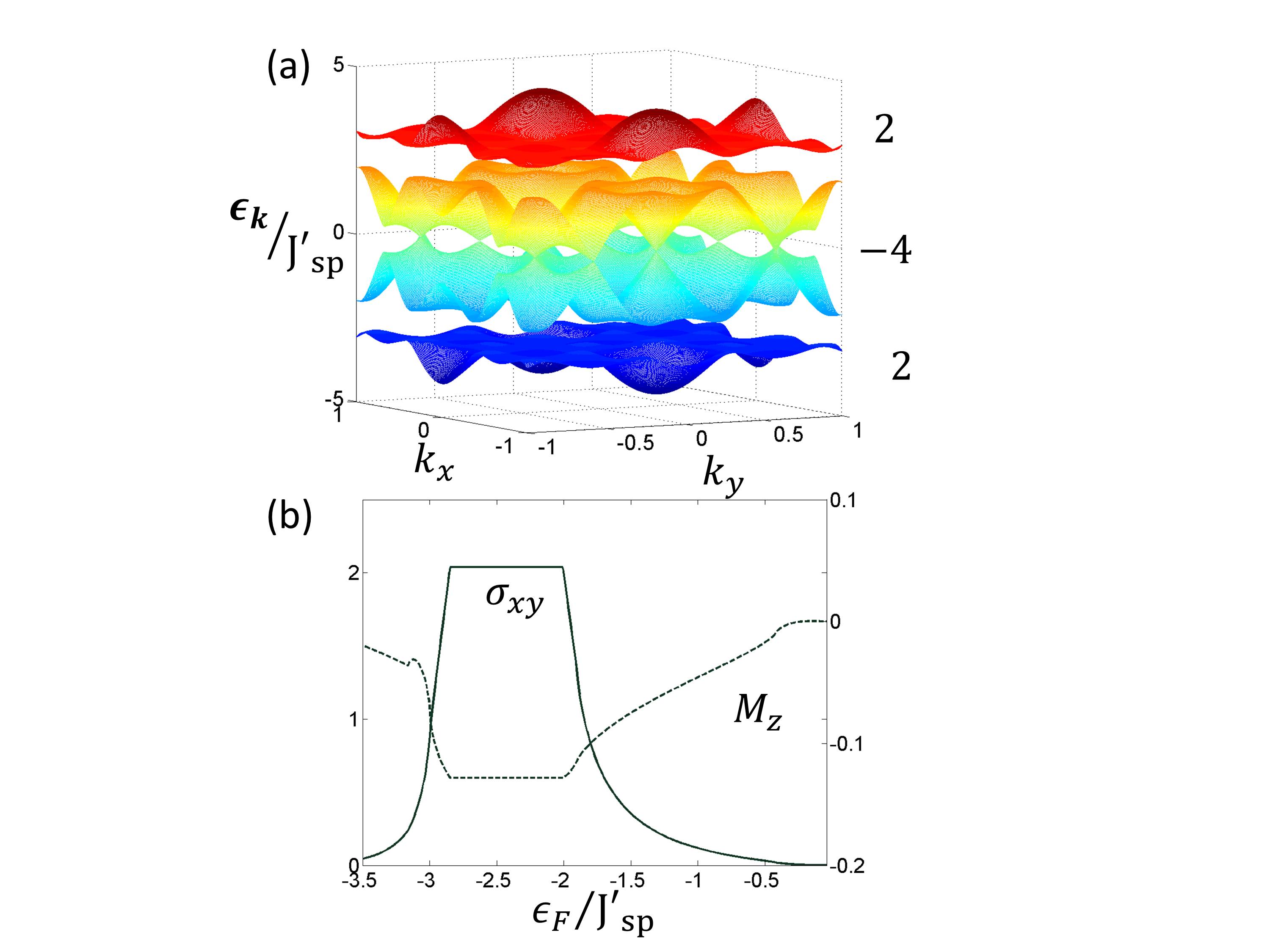}
\caption{\label{fig2} (a) The dispersion relation for  $\Phi=0$ which corresponds to staggered flux with $\varphi_1=\varphi_2=2\pi/3$ and $N=1$. The numbers on the right hand side of the figure show the invariant Chern numbers corresponding to the respective bands. {As we have seen in Fig.\ref{fig1}(a), there are multiple Dirac cones which touch each other in the zero flux limit. Thus, when there is a band gap, there are contributions to Berry curvature coming from all such points. This gives rise to bands with large Chern numbers $(2,-4,2)$}. Plot (b) represents the Hall conductivity $\sigma_{xy}$ and the magnetization $\mathcal{M}_z$ for that case.}
\end{figure}

 Finally, consider  the strong flux limit, e.g the case of  $\Phi=0$ where the flux through each plaquette is $2\pi/3$. Strong flux results in lifting the degeneracy between the first two bands (Fig.~\ref{fig2}, top plot). The middle two bands still touch each other in the form of Dirac cones. With the degeneracy lifted, one can define Chern numbers given by $\nu=(2,-4, 2)$ resulting in the appearance of Quantum Anomalous Hall effect. We have also calculated the Hall conductivity and when the Fermi energy of the excess fermions 
resides in the band gap, conductivity becomes integer valued (Fig.~\ref{fig2}, bottom plot). The magnitude of the band gap is $\approx J_{\rm sp}$. For a triangular lattice (lattice constant $a=500$nm)with lattice depth of $6E_R$ and transverse frequency of $10E_R$, the $sp$ tunneling strength in the harmonic approximation is given by $J_{\rm sp} \sim 0.008E_R$ assuming the scattering length of $-400$ Bohr radius.  This corresponds to a band gap {of about} $\sim 10$ nano-Kelvin which determines the temperature regime  where the Hall phase can be observed. For the Dice lattice with dilute impurities, the Hall conductivity presented in this paper remain unchanged due to the topological nature of the Berry curvature for the dispersion bands \cite{Nagaosa}. The band topology discussed here can be measured in principle by using recently proposed methods of Ramsey interferometry and Bloch oscillations \cite{Dem1, Dem2}, or from momentum distribution from Time-of-Flight images \cite{Duan}. Moreover, the generation 
of local orbital angular momentum due to broken time-reversal symmetry in the chiral $p$-orbitals can also be detected by time-of-flight measurements \cite{Hem}.

\section*{Conclusions} To summarize, we considered an unequal mixture of attractively interacting fermions in a shaken triangular lattice. Pairing produces immobile composites that gives rise to Dice lattice for the excess fermions. Adjustments of shaking frequency and amplitude allow to make intra-band tunnelings negligible while resonantly enhancing interaction-induced $sp$-tunnelings for the excess fermions. Moreover, shaking leads to the controlled staggered magnetic field and induces (on the $p$-orbitals) non-Abelian character of the system. Their joint effect leads to spontaneous chiral magnetization (due to time reversal symmetry breaking) along with appearance of Anomalous Hall effect.
Many fascinating question related to the findings here can be investigated further including the role of impurities, long-range interaction etc. Moreover, by using dipolar atoms, one can further study many-body effects like superconductivity \cite{Lim2, Pietro}, density-waves in presence of the artificial non-Abelian gauge fields
presented here.

\section*{Acknowledgments}
We thank M. Lewenstein and K. Sacha for enlightening discussions.
This work was realized under National Science Center (Poland) project No. DEC-2012/04/A/ST2/00088. A.P. is supported by the International PhD Project "Physics of future quantum-based information technologies", grant  MPD/2009-3/4 from Foundation for Polish Science and by the University of Gdansk grant BW 538-5400-B169-13-1E.

\section*{Author Contributions}
O.D., A.P., J.Z. conceived the idea, performed derivations and calculations, discussed the results and wrote the manuscript.

\section*{Additional Information}
The authors declare no competing financial interests.

\section*{Methods}
\label{methods}
\subsection*{The model Hamiltonian}

We consider an unequal mixture of two-species ultracold fermions (denoted by $\uparrow,\downarrow$) assuming strong attractive interactions between two species. It is then energetically favorable for fermions to pair, the low energy system is then effectively composed of paired composites and the excess $\da$ fermions.  We denote the creation and annihilation operators for $\ua$ fermions as $  \hat{s}^{\dagger}_{\ua\mathbf{i}}$ and $  \hat{s}^{}_{\ua\mathbf{i}}$. For the more abundant $\da$ fermions we
include both $s$ and $p$ orbitals denoting the corresponding operators as $\hat{s}^\dagger_{\da\mathbf{i}},\hat{s}_{\da\mathbf{i}}, \hat{p}^{\dagger}_{\da\pm\bm{i}}, \hat{p}^{}_{\da\pm\bm{i}}$.  In the main text, for  simplicity, we have neglected $\ua$-fermion tunneling and all the intra-band tunnelings for $\da$-fermions from the beginning. Here, let us derive the Hamiltonian without these assumptions and show that, indeed, these effects may be neglected. 

The full time-dependent Hamiltonian $H(t)$ consists of three parts  $H(t)=H_{\rm tun}+H_{\rm onsite}+H_{\rm shaking}$. The first, $H_{\rm tun}$ describes  the tunnelings, $H_{\rm onsite}$ describes the on-site interactions and $H_{\rm shaking}$ describes the shaking. Together they read:
\begin{eqnarray} \label{hamtotal}
H_{\rm tun}=&& J_{0}\sum_{\mathbf{i}, \bm{\delta}}{\hat s_{{\ua\mathbf{i}}}}^{\dagger}  \hat{s}^{}_{\ua\mathbf{i}+\bm{\delta}} + J_{0} \sum_{\mathbf{i}, \bm{\delta}}{\hat s_{\da\mathbf{i}}}^{\dagger}  \hat{s}^{}_{\da\mathbf{i}+\bm{\delta}} \nonumber\\
+ \sum_{\mathbf{i},\bm{\delta},\sigma}J_1^{\sigma\bm{\delta}}&&\hat{p}^{\dagger}_{\da\sigma,\mathbf{i}} \hat{p}^{}_{\da\sigma,  \mathbf{i}+\bm{\delta}} + \sum_{\mathbf{i},\bm{\delta},\sigma}J_{11}^{\sigma\bm{\delta}} \hat{p}^{\dagger}_{\da\sigma,\mathbf{i}}(\hat{n}^\ua_{\mathbf{i}}+\hat{n}^\ua_{\mathbf{i}+\bm{\delta}}) \hat{p}^{}_{\da\sigma,  \mathbf{i}+\bm{\delta}}
\nonumber \\ 
+&& \frac{J_{sp}}{\sqrt{2}} \sum_{\mathbf{i},\bm{\delta},\sigma=\pm} \left (f_{\bm{\delta}} \hat{p}^{\dagger}_{\da\sigma,\mathbf{i}}\hat{n}^\ua_{\mathbf{i}} \hat{s}^{}_{\da\mathbf{i}+\bm{\delta}} 
+{\it h.c} \right )
, \\
H_{\rm onsite}=&& U_{2}\sum_{\mathbf{i}}{\hat n_{ \mathbf{i}}}^{\ua}{\hat n_{s\mathbf{i}}}^{\da} + U_{01}\sum_{\mathbf{i}, \sigma=\pm}{\hat n}^{\da}_{\sigma\mathbf{i}}{\hat n}_i^\ua+E_1 \sum_{\mathbf{i},\sigma=\pm}{\hat n_{\sigma\mathbf{i}}}^{\da}, \nonumber\\
H_{\rm shaking} =&& \sum_{\mathbf{i}} \mathbf{i}.\mathbf{F}_t (\hat{n}^{\da}_{s\mathbf{i}}+\hat{n}^{\da}_{+\mathbf{i}}+\hat{n}^{\da}_{-\mathbf{i}}+\hat{n}^{\ua}_{s\mathbf{i}}).\nonumber
\end{eqnarray}
Here, ${\hat n}_i^{\ua(\da)}$, denote number operators of $\ua$ ($\da$) fermions respectively while $\hat{n}^{\da}_{\pm\mathbf{i}}$ are number operators for the $\da$ $p$-fermions with $\pm$-chirality. The same amplitude, $J_0$ corresponds to the standard tunneling between $s$ orbitals, the corresponding tunneling in the $p$-band  is described by $J_{1}^{\sigma\bm{\delta}}$. Moreover, we include density induced (bond-charge)  intra-band tunneling for $p$-orbitals with strength $J_{11}^{\sigma\bm{\delta}}$. ${J}_{sp}$ is the amplitude of the hopping between $s$ and $p$ bands which is also induced by the interaction with $\ua$ fermions. The various tunneling processes in Hamiltonian \eqref{hamtotal} are shown in Fig. \ref{sup_fig}. 
The tunneling amplitudes are given by
\begin{eqnarray}
J_0 &=& \int\int \mathcal{W}^{00}_{\mathbf{i}}(x,y)H_{\rm latt}\mathcal{W}^{00}_{\mathbf{i}+\bm{\delta}}(x,y)dxdy \nonumber\\
J_1^{\sigma\bm{\delta}} &=& \int\int [\mathcal{W}^{\sigma}_{\mathbf{i}}(x,y)]^{*}H_{\rm latt}\mathcal{W}^{\sigma}_{\mathbf{i}+\bm{\delta}}(x,y)dxdy \\
J_{11}^{\sigma\bm{\delta}} &=& g_{\rm 2D} \int\int [\mathcal{W}^{\sigma}_{\mathbf{i}}(x,y)]^{*}[\mathcal{W}^{00}_{\mathbf{i}}(x,y)]^2\mathcal{W}^{\sigma}_{\mathbf{i}+\bm{\delta}}(x,y)dxdy,\nonumber
\end{eqnarray}
where $\mathcal{W}^{00}_{\mathbf{i}}(x,y)$ is the Wannier function of the $s$-band and $\mathcal{W}^{\sigma}_{\mathbf{i}}(x,y)$ with $\sigma=\pm$ are the Wannier functions corresponding to $p_{+}$-and $p_{-}$-bands in the harmonic approximation for the triangular lattice potential. The single particle Hamiltonian for the triangular lattice is denoted by $H_{\rm latt}$. 

{Note that the Hamiltonian (\ref{hamtotal}) does not contain tunnelings of the composites themselves. Such a pair tunneling term can arise due to interaction \cite{Dutta2015} but is 3-4 orders of magnitude smaller than other tunneling terms. Thecomposites can also tunnel via higher-order processes (discussed in \cite{Strohmaier}). The leading term of this collective tunneling is of the second order  \cite{Mic} with the corresponding amplitude being proportional to $J_0^2/|U_2|$, i.e. very small assuming strong attraction. The effect is further reduced by assumed shaking - modification of effective $J_0$ - so such tunnelings can be safely neglected.}

\subsection*{Low-energy and resonant subspaces}

Now we define the low-energy subspace and the resonant subspace which are coupled by the driving (shaking). First we assume the strong interaction limit i.e. $J_0, J_1^{\sigma\bm{\delta}}, J_{11}^{\sigma\bm{\delta}}, J_{sp} \ll U_2, U_{01}$. Yet larger energy scale is set by single particle energy of the $p$ band $E_1$ and the shaking frequency. Thus we assume  $U_2, U_{sp}\ll E_1 \sim \Omega$. $|U_2|$ - the strength of attraction between $\ua$ and $\da$ fermions sets the low-energy scale, thus we restrict the analysis to the subspace of Hilbert space where all $\ua$ minority fermions are paired with their $\da$ partners. Thus the low-energy local subspace is spanned by  ${\hat s_{\da{\mathbf{i}}}}^{\dagger}{\hat s_{\ua{\mathbf{i}}}}^{\dagger}|0\rangle, {\hat s_{\da{\mathbf{i}}}}^{\dagger}|0\rangle $ states. As we will show below, due to the $sp$ tunneling and periodic driving this subspace is resonantly connected to the subspace where a paired site can be occupied by $p$-orbital fermions, $ \hat{p}^{\dagger}
_{\da\sigma,\mathbf{i}}{\hat s_{\da{\mathbf{i}}}}^{\dagger}{\hat s_{\ua{\mathbf{i}}}}^{\dagger}|0\rangle$ with energy $E_1+U_{01}$. Therefore, from now on our Hilbert space will consists of  ${\hat s_{\da{\mathbf{i}}}}^{\dagger}{\hat s_{\ua{\mathbf{i}}}}^{\dagger}|0\rangle, {\hat s_{\da{\mathbf{i}}}}^{\dagger}|0\rangle, \hat{p}^{\dagger}_{\da\sigma,\mathbf{i}}{\hat s_{\da{\mathbf{i}}}}^{\dagger}{\hat s_{\ua{\mathbf{i}}}}^{\dagger}|0\rangle $ states.

We now apply the unitary transformation, $\hat{U}_t=\exp [- { i} H_{\rm onsite} t - { i} \int^t_0 H_{\rm shaking}(t')dt']$ transferring the time-dependence in  the total Hamiltonian $H(t)$ into the tunneling amplitudes. The new Hamiltonian 
$H'=\hat{U}^\dagger H \hat{U} - {i}\hat{U}^\dagger [d_t \hat{U}]$ is given by
\begin{eqnarray} \label{timeham}
H'&=& J_{0} \sum_{\mathbf{i}, \bm{\delta}} \exp[-iU_{2}(\hat{n}^\da_{\mathbf{i}}-\hat{n}^\da_{\mathbf{i}+\bm{\delta}})t-i \bm{\delta}\cdot\mathbf{W}_t] {\hat s_{\ua{\mathbf{i}}}}^{\dagger}  \hat{s}^{}_{\ua\mathbf{i}+\bm{\delta}}
\nonumber\\
&+&J_{0} \sum_{\mathbf{i}, \bm{\delta}} \exp[-iU_{2}(\hat{n}^\ua_{\mathbf{i}}-\hat{n}^\ua_{\mathbf{i}+\bm{\delta}})t-i \bm{\delta}\cdot\mathbf{W}_t] {\hat s_{\da{\mathbf{i}}}}^{\dagger}  \hat{s}^{}_{\da\mathbf{i}+\bm{\delta}}\nonumber\\
 &+& \sum_{\mathbf{i},\bm{\delta},\sigma}\exp[-iU_{01}(\hat{n}^\ua_{\mathbf{i}}-\hat{n}^\ua_{\mathbf{i}+\bm{\delta}})t-i \bm{\delta}\cdot\mathbf{W}_t] \nonumber \\
&\times & \hat{p}^{\dagger}_{\da\sigma,\mathbf{i}} \left [ J_1^{\sigma\bm{\delta}} +  J_{11}^{\sigma\bm{\delta}}(\hat{n}^\ua_{\mathbf{i}}+\hat{n}^\ua_{\mathbf{i}+\bm{\delta}}) \right ]  \hat{p}^{}_{\da\sigma, \mathbf{i}+\bm{\delta}}\\
&+& \frac{J_{sp}}{\sqrt{2}}\sum_{\mathbf{i},\bm{\delta},\sigma=\pm} \exp[-i E_1 t - i U_{01} t-i \bm{\delta}\cdot\mathbf{W}_t]f_{\bm{\delta}} \hat{p}^{\dagger}_{\da\sigma,\mathbf{i}}\hat{n}^\ua_{\mathbf{i}} \hat{s}^{}_{\da\mathbf{i}+\bm{\delta}} ,\nonumber 
\end{eqnarray}
where $\mathbf{W}_t=\int^t_0 \mathbf{F}_{t'}dt'$. We expand the exponential functions in \eqref{timeham} as:
$\exp[-i \bm{\delta}\cdot\mathbf{W}_t]=\sum_n \mathcal{J}_n(K_{\bm{\delta}}/\Omega) \exp \left[-i n \Omega t \right]$. Then as $U_2 \ll \Omega$, after rotating-wave approximation (RWA) and projecting on our local Hilbert space, the first term of Hamiltonian \eqref{timeham} may be resonant only if $U_2$ contribution vanishes. Since this term corresponds to $\ua$-fermion tunneling (which appear  paired only in our subspace) this process is possible only if a paired state and a $\da$-fermion in $s$-orbital are neighbors (Fig. 1(a)). Otherwise  the pair (composite) is pinned.  Similarly, the second term may be resonant (${n}^\ua_{s\mathbf{i}}={n}^\ua_{s\mathbf{i}+\bm{\delta}}=0$) when a $\da$-fermion in $s$-orbital tunnels to a neighboring empty site. The third term gives a resonant contribution via the tunneling process depicted in Fig.\ref{sup_fig}(b). 
After RWA, all  the time-independent tunneling amplitudes 
of the above intra-band tunnelings are changed by a factor $\mathcal{J}_0(K_{\bm{\delta}}/\Omega)$. We see that to minimize the $ss$ and $pp$ tunnelings we have to tune the shaking amplitude such that $\mathcal{J}_0(K_{\bm{\delta}_0}/\Omega)=0$ and $\mathcal{J}_0(K_{\bm{\delta}_{-}}/\Omega)=0$. 
This assures that for the shaking phase $\Phi=0$, there is no intra-band tunneling along the $\bm{\delta}_{+}$ direction as $K_{\bm{\delta}_{+}}=K_{\bm{\delta}_{-}}$. 

In the last term of Hamiltonian \eqref{timeham}  the fast oscillation with $E_1+U_{01}$ frequency must be compensated by appropriate Fourier component yielding  the $sp$ resonant condition $E_1+U_{01}=N\Omega$. Inspecting the tunneling term  we see that, the tunneling in $p$-band is resonantly enhanced only when the composite density in neighboring sites $\mathbf{i}$ and $\mathbf{i}+\bm{\delta}$ follows the relation $({n}^\ua_{\mathbf{i}}-{n}^\ua_{\mathbf{i}+\bm{\delta}})=1$.
 Due to the type of $sp$ coupling in Hamiltonian, \eqref{timeham}, $p$-fermions may appear only in composite occupied sites.  
 This may occur only from a site occupied by a lonely $\da$-fermion (if there were a composite at that site, an additional energy difference, $U_2$, the pair energy would appear bringing the system out of the chosen resonance). After carrying RWA and in the limit of vanishing intra-band tunneling, the effective Hamiltonian reads,
\begin{eqnarray}\label{avgtime}
H' &=& \frac{J_{sp}}{\sqrt{2}}\sum_{\mathbf{i},\bm{\delta},\sigma=\pm} \mathcal{J}_N(K_{\bm{\delta}}/\Omega) \exp\left[-i\sigma\tan^{-1}\left(\delta_{y}/\delta_{ x}\right)\right]\hat{p}^{\dagger}_{\downarrow \sigma,\mathbf{i}}\hat{n}^\ua_{\mathbf{i}} \hat{s}^{}_{\da \mathbf{i}+\bm{\delta}}, 
\end{eqnarray}
where $\mathcal{J}_N(x)$ defines Bessel function of order-$N$. We see that, one can control the different tunneling amplitudes by tuning the shaking amplitude, frequency and interaction strength.

When the shaking phase $\Phi \neq 0$, along $\bm{\delta}_{0}$ and $\bm{\delta}_{-}$ directions the intra-band tunneling still vanishes, but remains nonzero along $\bm{\delta}_{+}$ direction. 
Amplitude of the latter can be tuned to values smaller than the $sp$-tunneling amplitude by changing the interaction strength. Moreover, once the Dice structure of the composites is created, the only possible tunneling along ${\bm{\delta}_{+}}$ direction is the inter-band $sp$ tunneling (compare Fig.1(a) in the main text). So, adding small intra-band tunneling due to a finite shaking phase will not destabilize the Dice structure.

\subsection*{Effects of tunneling on the emergent lattice}

\begin{figure}
\includegraphics[width=140mm,natwidth=610,natheight=642]{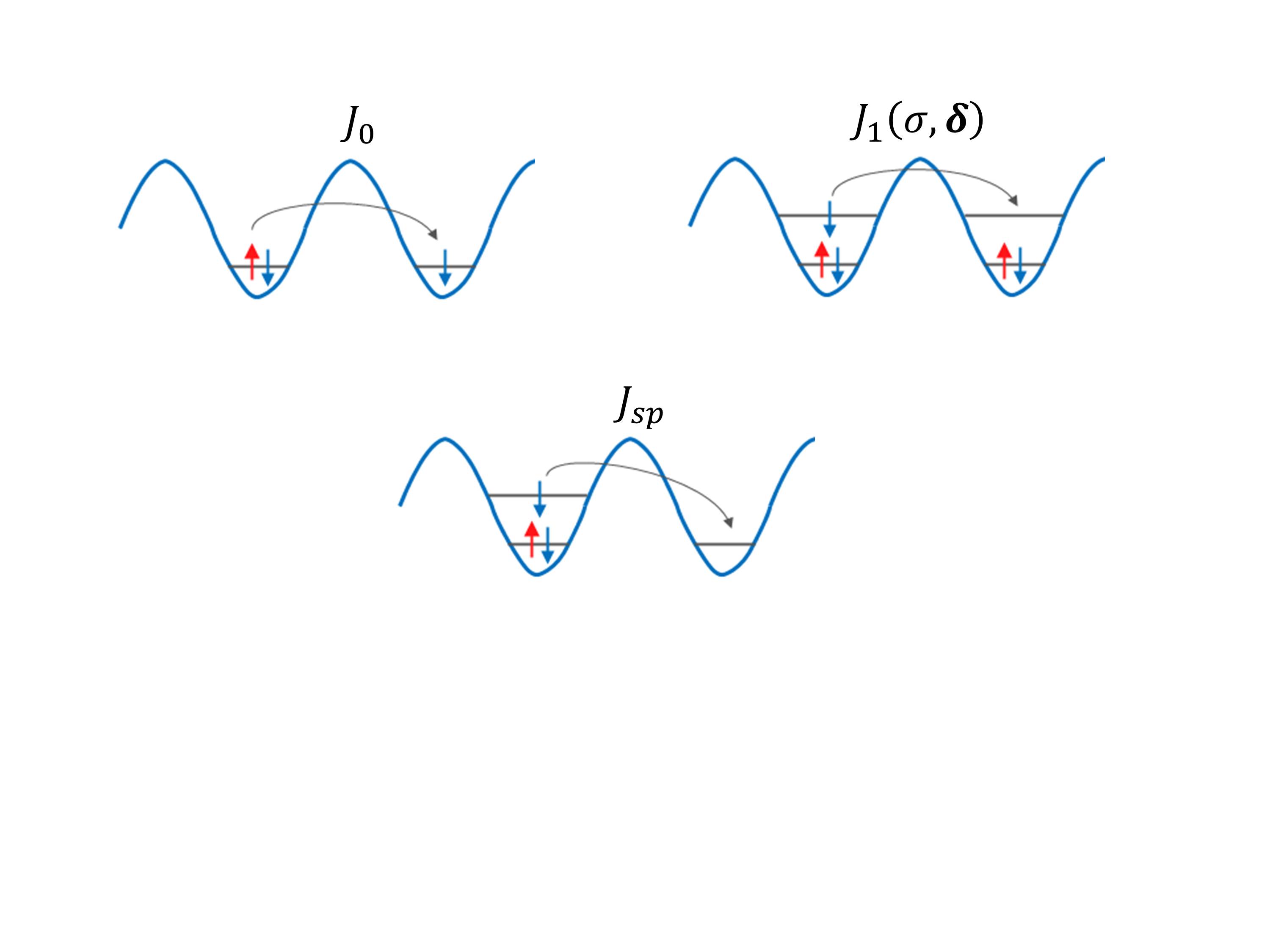}
\vspace{-2.0cm}
\caption{\label{sup_fig} Pictorial representation of different tunneling processes in Hamiltonian, Eq.\eqref{hamtotal}. The top left panel describes the resonant tunneling process where a $\uparrow$ fermion from a composite tunnel to a neighboring excess $\downarrow$-fermion occupied site.  This tunneling process corresponds to the first term in Hamiltonian $H_{\rm tun}$. The top right panel describes the process when an excess fermion can tunnel in the $p$-band resonantly provided both the sites are already occupied by composites. This reflects the third term in Hamiltonian $H_{\rm tun}$. The bottom figure depicts the interaction induced $sp$ tunneling amplitude.}
\end{figure}

{In this section, we discuss the effect of the tunneling on the Dice lattice structure. As discussed before,  a composite can tunnel to a vacant site only via  higher order processes \cite{Strohmaier,Mic} which are negligible for large $|U_2|$. So the only way a composite can tunnel is if the minority fermion  tunnels to a site already occupied by a majority fermion in $s$-orbital site as shown in the first figure in Fig.\ref{sup_fig}. Such a process  can be described by an effective tunneling for the composite coupled to the tunneling of the excess fermions in the opposite direction. To investigate the effect of such a tunneling we use a one-dimensional minimal model,
\begin{eqnarray}
H_{\rm min}&=& -\frac{J_{\rm sp}}{\sqrt{2}}\sum_{\langle ij \rangle} \left[\hat{p}^{\dagger}_{i}\hat{n}^c_{{i}} \hat{s}_{{j}} + {\it h.c}\right] - J_0 \sum_{\langle ij \rangle} \left[ \hat{c}^{\dagger}_{i}\hat{c}_{{j}} \hat{s}^{\dagger}_{{j}}\hat{s}_{{i}} \right],
\end{eqnarray}
where at each site $i$ we have only $s$- and $p$-orbitals, and $\langle ij \rangle $ denotes the nearest neighbors. We have introduced operators $\hat{c}_i, \hat{c}^{\dagger}_{i}$ as the composite annihilation and creation operators. The first term denotes the composite density dependent $sp$ tunneling of the excess fermions and the last term just denotes the composite tunneling and excess fermion tunneling. When $J_0=0$, the ground state is given by the composite structure, 
$n^c_{2i}=1, n^c_{2i+1}=0$ when composite filling is $n^c=1/2$. Such a density wave structure is equivalent to the Dice lattice structure we study in a triangular lattice. Due to the hardcore bosonic nature of the composites, we use a factorized variational composite wavefunction, 
$ \left | \Phi_c \right \rangle = \Pi_i \left |\Phi_c \right\rangle_i$, where $\left |\Phi_c \right\rangle_{2i} = \cos\theta |1\rangle_c + \sin\theta |0\rangle_c $ and $\left |\Phi_c \right\rangle_{2i+1} = \cos\theta |0\rangle_c + \sin\theta |1\rangle_c $ and $|1\rangle_c, |0\rangle_c$ denote a composite occupied or empty site. In the composite wavefunction ansatz, $\theta$ is the variational parameter. The density wave state at $J_0=0$ is obtained for $\theta=0$. Using such an ansatz, we can integrate over the composite subspace and get an effective Hamiltonian,
\begin{eqnarray}
H_{\rm eff}&=& \frac{J_{\rm sp}\cos^2\theta}{\sqrt{2}}\sum_{\langle i \rangle} \left[\hat{p}^{\dagger}_{2i}\hat{s}_{2i+1} + {\it h.c}\right] + \frac{J_{\rm sp}\sin^2\theta}{\sqrt{2}}\sum_{\langle i \rangle} \left[\hat{p}^{\dagger}_{2i+1}\hat{s}_{2i} + {\it h.c}\right] - J_0 \frac{\sin^2 2\theta}{4} \sum_{\langle ij \rangle} \hat{s}^{\dagger}_{{i}}\hat{s}_{{j}}. \nonumber\\
\end{eqnarray}
}
{Then we write the energy for excess fermion filling $n=1/4$ (this is $1/2$ of the previous value due to the doubling of number of degrees of freedom) for $\theta \ll 1$ and $J_{\rm sp} \gg J_0$ as $E_{\rm var}=\frac{2\sqrt{2}J_{\rm sp}}{\pi}\left[-1+\theta^2 \right] + O(\theta^4)$, which is independent of composite tunneling. From that we conclude that the energy is minimized for $\theta = 0$. For larger tunneling strength $J_0$, we have compared the energy of the homogenous state with $\theta=\pi/4$ and the density wave state with $\theta=0$ finding that the density wave state has lower energy as long as $J_{\rm sp} > 3J_0/4$. Though the present calculation is one-dimensional, the essential physics also applies to the more complicated situation of triangular lattice, where we expect the Dice lattice density wave structure to be stable even in the presence of small composite tunneling.}

\end{document}